\documentclass[aps,pre,nofootinbib,showpacs,superscriptaddress,twocolumn]{revtex4-2}
\usepackage[utf8]{inputenc}
\usepackage[T1]{fontenc}
\usepackage[english]{babel}
\usepackage{graphicx}
\usepackage{epsfig}
\usepackage{epstopdf}
\usepackage{amsmath}
\usepackage{amssymb}
\usepackage{times}
\usepackage{setspace}
\usepackage{verbatim}
\usepackage{color}
\usepackage{mathrsfs}
\usepackage{euscript}
\usepackage{mathptmx}

\usepackage{hyperref}
\hypersetup{
    colorlinks=true,
    linkcolor=blue,
    filecolor=magenta,      
    urlcolor=red,
    citecolor=red, 
    pdftitle={LR-FPUT},
    }
\begin{document}
\title{Heat transport in long-ranged anharmonic oscillator models}
\author{Debarshee Bagchi}
\email[E-mail address:]{debarshee.bagchi@icts.res.in}
\affiliation{International Centre for Theoretical Sciences, Tata Institute of Fundamental Research, Bengaluru, India}
\date{\today}

\begin{abstract}
In this work, we perform a detailed study of heat transport in one dimensional long-ranged anharmonic oscillator systems, such as the long-ranged Fermi-Pasta-Ulam-Tsingou model. For these systems, the long-ranged anharmonic potential decays with distance as a power-law, controlled by an exponent $\delta \geq 0$. For such a non-integrable model, one of the recent results that has captured quite some attention is the puzzling ballistic-like transport observed for $\delta = 2$, reminiscent of integrable systems. Here, we first employ the reverse nonequilibrium molecular dynamics simulations to look closely at the $\delta = 2$ transport in three long-ranged models, and point out a few problematic issues with this simulation method. Next, we examine the process of energy relaxation, and find that relaxation can be appreciably slow for $\delta = 2$ in some situations. We invoke the concept of nonlinear localized modes of excitation, also known as discrete breathers, and demonstrate that the slow relaxation and the ballistic-like transport properties can be consistently explained in terms of a novel depinning of the discrete breathers that makes them highly mobile at $\delta = 2$. Finally, in the presence of quartic pinning potentials we find that the long-ranged model exhibits Fourier (diffusive) transport at $\delta = 2$, as one would expect from short-ranged interacting systems with broken momentum conservation. Such a diffusive regime is not observed for harmonic pinning.

\end{abstract}
\maketitle

\section{Introduction}
Heat transport is an unsolved paradigmatic problem in non-equilibrium statistical physics that has been actively investigated in the last few decades, both from theoretical and experimental perspectives \cite{dhar2008heat,lepri2003thermal,prosen2005normal,liu2012anomalous,benenti2020anomalous}. Arguably, the most surprising result that has been obtained from these studies is that many one dimensional ($1D$) systems do not obey the celebrated Fourier's law of heat conduction, $j = -\kappa \frac{dT}{dx}$, where $j$ is the thermal flux and $dT/dx$ is the thermal gradient. This implies that thermal conductivity $\kappa$ is ill-defined in the thermodynamic limit $N \to \infty$, and $\kappa$ for finite systems scales algebraically with the system size $N$ as $\kappa \sim N^\alpha$, where $0 < \alpha \le 1$. Apart from a few exceptions, thermal transport is generally found to be anomalous (i.e., non-Fourier, $\alpha > 0$) for momentum conserving $1D$ system, and diffusive (i.e., Fourier, $\alpha = 0$) in the presence of external forces (momentum non-conservation). For integrable systems, thermal transport is known to be ballistic in nature -- thermal current $j$ is independent of $N$, implying $\alpha = 1$, and temperature profiles are flat. Although a lot still remains to be proved rigorously, nonetheless we have already gathered a comprehensive general understanding of heat transport in $1D$ systems, most particularly for models with nearest-neighbor interactions.

Quite recently the problem of heat transport has been extended to lattice models with long-ranged interactions (LRIs) \cite{pereira2013increasing,avila2015length,olivares2016role,bagchi2017thermal,bagchi2017energy,iubini2018heat,wang2020thermal,tamaki2020energy,livi2020heat}. In this work, LRIs are defined in terms of the pair potential $V(d_{ij})$ that decays with the distance $d_{ij}$, between the $i-$th and the $j-$th particle, as $V(d_{ij}) \sim d_{ij}^{-\delta}$, where $\delta \geq 0$. The parameter $\delta$ can be tuned to manipulate the range of the long-ranged interactions: thus $\delta \to \infty$ corresponds to nearest-neighbor interactions, whereas $\delta = 0$ is the mean field scenario. These $1D$ systems are non-additive for $0 \leq \delta \leq 1$ and additive for $\delta > 1$. For extensive reviews on LRIs, see Refs. \cite{dauxois2002dynamics,campa2009statistical,bouchet2010thermodynamics,levin2014nonequilibrium}.
LRIs are ubiquitous at all length-scales in nature, from gravitation between celestial objects \cite{padmanabhan1990statistical} to Coulomb interaction between charged particles in nanosystems \cite{french2010long}, but their properties are not well understood yet.
Systems with LRIs often exhibit dynamic and thermodynamic properties that are drastically different from that in the conventional well-behaved short-ranged systems -- a few such properties are breakdown of ergodicity, inequivalence of statistical ensembles, negative specific heat, violation of zeroth law of thermodynamics, extremely long-lived quasi-stationary states. In the following, we highlight some recent works concerning heat transport in $1D$ long-ranged models that are relevant for this study.

In one of the works, heat transport in the long-ranged version of the planar rotor model was investigated using nonequilibrium molecular dynamics (NEMD) simulations, and it was found that there are two distinct transport regimes  \cite{olivares2016role}. In the thermodynamics limit, for $0 \leq \delta \lesssim 1$, the long-ranged rotor model is a thermal insulator ($\kappa \to 0$ as $N \to \infty$), whereas, for $\delta \gtrsim 1$, one finds a thermal conductor obeying Fourier's law ($\kappa \sim N^0$). 
In another work, using NEMD and equilibrium simulations, the long-ranged Fermi-Pasta-Ulam-Tsingou (LR-FPUT) model was studied and a strikingly different transport behavior has been observed \cite{bagchi2017thermal}. For the LR-FPUT model, $\kappa$ is non-monotonic with the range parameter $\delta$ and has a maximum at $\delta = 2$. Thus, very long-ranged and very short-ranged systems have a low conductivity, whereas, for an optimum value $\delta = 2$ one obtains maximum thermal conductivity. Moreover, it is found that for all $\delta \neq 2$ thermal transport is anomalous (super-diffusive), but at the ``special point'' $\delta = 2$ thermal transport seems to have a ballistic-like nature (we find $\kappa \sim N$ and almost flat temperature profiles). Similar $\delta=2$ transport properties have also been reported for a few other variations of the LR-FPUT model \cite{iubini2018heat}, indicating that the observed properties are quite general and robust, although doubts were raised about finite-size effects and on the possibility of quasi-integrability.

Recently, this problem has been studied again for a similar $1D$ long-ranged anharmonic oscillator model using the reverse nonequilibrium molecular dynamics (RNEMD) simulations, and contrary to previous results, it was suggested that transport for $\delta = 2$ is not ballistic-like ($\alpha \approx 1$) but super-diffusive in nature with an exponent $\alpha \approx 0.71$ \cite{wang2020thermal}. Note that this is not only a significant quantitative discrepancy but a puzzling qualitative contradiction as well. In Ref. \cite{tamaki2020energy}, another long-ranged lattice model was analytical studied and one of the main results reported in this work is that anomalous (non-Fourier) heat transport is possible even in the presence of momentum nonconservation. 

In this article, we revisit some of these issues, with the intention of resolving the contradiction related to the $\delta = 2$ transport, and to better understand the heat transport properties in a general class of $1D$ anharmonic oscillator models with long-ranged interactions. The remainder of the paper is organized as follows.
In Sec \ref{model}, we specify the LR-FPUT model, which is the main focus of this work, along with another related model, and some details of the numerical techniques that have been used.
In Sec. \ref{rnemd}, we perform RNEMD simulations to understand why one gets qualitatively different transport behavior from NEMD and RNEMD simulations.
Next, in Sec. \ref{Erelax}, we look at the process of energy relaxation in the LR-FPUT model to highlight the uniqueness of the $\delta = 2$ point.
In Sec. \ref{DB}, we investigate the origin of the intriguing heat transport properties of the LR-FPUT model by examining the dynamics of discrete breathers in a few models, using {\it boundary cooling} experiments.
Finally, in Sec. \ref{pinning}, we study heat transport with the inclusion of external pinning potentials, in order to find out if one can observe Fourier behavior in the LR-FPUT system at $\delta=2$.
We conclude this work in Sec. \ref{conclusion} with a short summary and a discussion.

\section{Models and Methods}
\label{model}
In its most general form, the Hamiltonian $\mathcal{H}$ for this class of long-ranged anharmonic oscillator models can be expressed as
\begin{equation}
 \mathcal{H} = \sum_i \Big\{ \frac 12 p_i^2 + V_{SR}(x_{i+1} - x_{i}) + \sum_{j} V_{LR}(x_j - x_i) + U(x_i) \Big\},
 \label{eq:LR-FPUT}
\end{equation}
where all the $N$ particles have mass $m_i = 1$ ($1 \leq i,j \leq N$). Here $\{p_i, x_i\}$ represent the momentum and the displacement (from mean position) of the $i-$th particle on the $1D$ lattice. $V_{SR}(x)$ and $V_{LR}(x)$ represent the short-ranged (nearest-neighbor) and the long-ranged part of the pairwise interaction potential respectively. One can also include an external onsite (pinning) potential denoted by $U(x)$. Following our previous work \cite{bagchi2017thermal}, a majority of the results that we present here are for the LR-FPUT model that was introduced first in Ref. \cite{bagchi2016sensitivity}, described by the Hamiltonian
\begin{multline}
\textbf{LR-FPUT:} \\
\mathscr{H}_1 = \sum_{i}\frac 12 p_i^2 +  \sum_i \frac k2 (x_{i+1} - x_{i})^2 + \frac{\beta}{4\widetilde{N}} \sum_{i,j>i} \frac {(x_j - x_i)^4}{d_{ij}^{\delta}}.
\label{eq:lrfput}
\end{multline}
We set the spring constant $k = 1$, and also $\beta = 1$, without loss of generality. Here $d_{ij} = |i-j|$ is taken as the shortest distance between the mean positions of the oscillators $i$ and $j$. The scaling factor $\widetilde N = \frac 1N \sum_i \sum_j |i-j|^{-\delta}$ makes the total energy extensive for all values of $\delta$, and $U(x) = 0$. 

A variant of the LR-FPUT model that we have also studied in this work is what we refer to as the long-ranged {\it quartic} Fermi-Pasta-Ulam-Tsingou (LR-QFPUT) model, obtained by setting $k = 0$ in Eq. (\ref{eq:lrfput}). Written explicitly, the Hamiltonian for LR-QFPUT is
\begin{multline}
\textbf{LR-QFPUT:} ~~~ \mathcal{H}_2 = \sum_{i}\frac 12 p_i^2 + \frac{\beta}{4\widetilde{N}}\sum_{i,j>i} \frac{(x_j-x_i)^4}{d_{ij}^\delta}.
\label{eq:lrqfput} 
\end{multline}
This model does not have a short-ranged nearest-neighbor interaction term unlike the LR-FPUT model, and is therefore similar in this regard to the model discussed in Ref. \cite{iubini2018heat}. A few other variants of the LR-FPUT model will be mentioned at appropriate places in the following sections.

We have probed the transport properties of these models using a variety of equilibrium and nonequilibrium techniques. This requires us to study both isolated closed rings and open chains with boundary heat baths (denoted by temperatures $T_1$ and $T_2$ for the left and the right end). For closed rings, the conventional periodic boundary conditions are used, while for open chains we have implemented the fixed boundary conditions. For simulations with heat baths, the standard Langevin thermostat is used to maintain the desired temperature. For integrating the equations of motion, we have employed the velocity-Verlet algorithm with small time-step, typically in the range $0.005 \leq \Delta t \leq 0.10$, depending on the accuracy of energy conservation desired, and other system parameters. For the  initial conditions, the particle positions are chosen randomly from a uniform distribution, whereas the velocities are drawn randomly from a Gaussian distribution, both with zero mean. For microcanonical simulations in periodic systems, the center of mass velocity is set to zero, and the velocities are re-scaled to obtain the desired energy.

\section{RNEMD simulations}
\label{rnemd}
As mentioned earlier, in Ref. \cite{wang2020thermal}, the RNEMD simulation technique has been employed for studying heat transport. There, it has been suggested that the ballistic-like transport, obtained for the LR-FPUT model at $\delta = 2$ using NEMD simulations, appears because of strong finite-size effects arising due to the use of boundary heat baths and the $\widetilde{N}$ scaling of $V_{LR}(x)$ in Eq. \ref{eq:lrfput}. The exponent reported in Ref. \cite{wang2020thermal}, obtained by removing the $\widetilde{N}$ scaling and using RNEMD simulations, is $\alpha \approx 0.71$ at $\delta = 2$. Thus, it seems to be in direct contradiction with the NEMD results ($\alpha \approx 1$) that have been found consistently for a few different variants of the LR-FPUT model \cite{bagchi2017thermal, iubini2018heat}. In this section, we perform RNEMD simulations for LR-FPUT (and some related models) to find out the reasons for this discrepancy. This is important to understand, in order to reconcile the results that have been previously reported with the results obtained in this work.

The RNEMD method is a relatively new simulation technique that was proposed by M\"{u}ller-Plathe in $1997$ \cite{muller1997simple} as an alternative method for investigating transport phenomena. Originally, it was implemented for transport studies of Lennard-Jones fluids, but later on, it has been applied to more complex systems, such as, carbon and graphene nanosystems \cite{alaghemandi2009thermal, bagri2011thermal, zhang2012low, xu2014length,elapolu2019phononic,degirmenci2020reverse}. The RNEMD method has been found to be quite robust for studying thermal transport properties, but also has some issues that have been pointed out in recent works, and improvements have been suggested in some cases \cite{kuang2010gentler,tenney2010limitations,kuang2012velocity, yang2020effects}.

Unlike the usual NEMD simulations where one imposes a thermal gradient and measures the thermal current, in RNEMD simulations the relation between cause and effect is ``reversed''. In its simplest version, to perform RNEMD simulations, a microcanonical system (isolated ring) is subdivided into $w$ slabs (or bins) such that each slab contains a macroscopic number of particles $n$, but still smaller than the total number of particles $N$ in the system. One of these slabs, say the $w = 1$ slab, is labeled as the {\it cold} slab and another slab, the $w/2$ slab, is labeled as the {\it hot} slab. 

Starting from an equilibrated system, the particles are evolved in time obeying the usual Hamiltonian dynamics. However, every once in a while, the hottest particle (with velocity $v_h$) in the cold slab is swapped with the coldest particle (with velocity $v_c$) in the hot slab, provided $v_h^2 > v_c^2$. The time interval for performing this velocity swap is denoted by $\tau_{s}$. If one keeps on repeating these two operations -- Hamiltonian dynamics and velocity swapping -- the system eventually reaches a nonequilibrium steady state (NESS). At NESS, a steady {\it backward} current flows through the lattice from the hot bath to the cold bath, as a consequence of the ``artificial'' velocity swapping. The forward and the backward currents at NESS will be (statistically) equal and opposite to each other, and a temperature gradient will develop in both halves of the ring.

For RNEMD, the steady state heat current $j$ is precisely known and is computed as the net kinetic energy transported from the cold slab to the hot slab over all the velocity swap events: $j = \dfrac{\sum_{swaps} v_h^2 - v_c^2}{2 ~\#swaps}$. From the current $j$, the thermal conductivity can be obtained using Fourier's law: $\kappa = \frac{j}{dT/dx}$, where $\frac{dT}{dx}$ is computed numerically from the linear part of the temperature profile between the hot and the cold slab.

For our RNEMD simulations, we adopt a gentler approach. First, instead of exchanging the hottest and the coldest particles, we swap a moderately hot particle with a moderately cold particle (also suggested in Ref. \cite{muller1997simple}) and, second, we do it much less frequently, $\tau_s^{-1} = 0.02$, compared to $\tau_s^{-1} = 0.1$ in Ref.  \cite{wang2020thermal}. This is because, when the velocities of the hottest and the coldest particles are swapped, the temperature of the cold slab becomes very close to zero, and the temperature difference between the hot and cold slabs becomes very large, both of which we prefer to avoid in our simulations. Moreover, the aggressive swapping of velocities between particles of extreme energies may cause large thermal shocks in the system, leading to unwanted numerical artifacts. These issues can be avoided by making the swaps rare and only between moderately energetic particles. 

To perform the velocity swaps, we first sort (from hottest to the coldest) the $n$ particles in the hot and the cold slabs, and then we swap the $(n/2 - n_s + 1)-$th particle of the cold slab with the $(n/2 + n_s)-$th particle of the hot slab. Here $n_s$ is an integer and for most of our simulations, we set $n_s = 2$. (For $n_s = n/2$, we have velocity swaps between the hottest and the coldest particles). These modifications allow us to tune the temperatures of the two slabs, for $\delta=2$, to be approximately the same as the heat bath temperatures used for NEMD simulations in Refs. \cite{bagchi2017thermal,iubini2018heat}, $T_1 = 1.1$ and $T_2 = 0.9$ (see Fig. A.1a in Appendix). Note that the slab temperatures change as we alter $\delta$ and the parameters need to be re-calibrated in order to obtain the same hot and cold slab temperatures (see Fig. A.1b in Appendix).
\begin{figure}[htb]
{\includegraphics[width=4.5cm]{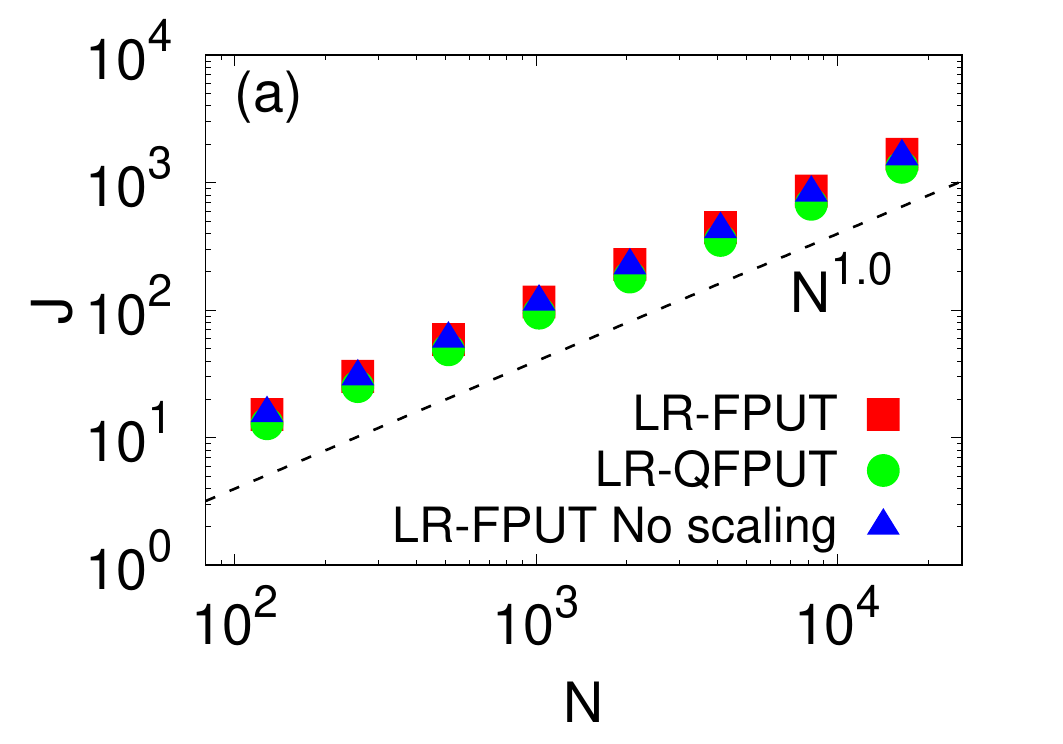}}\hskip-0.45cm
{\includegraphics[width=4.5cm]{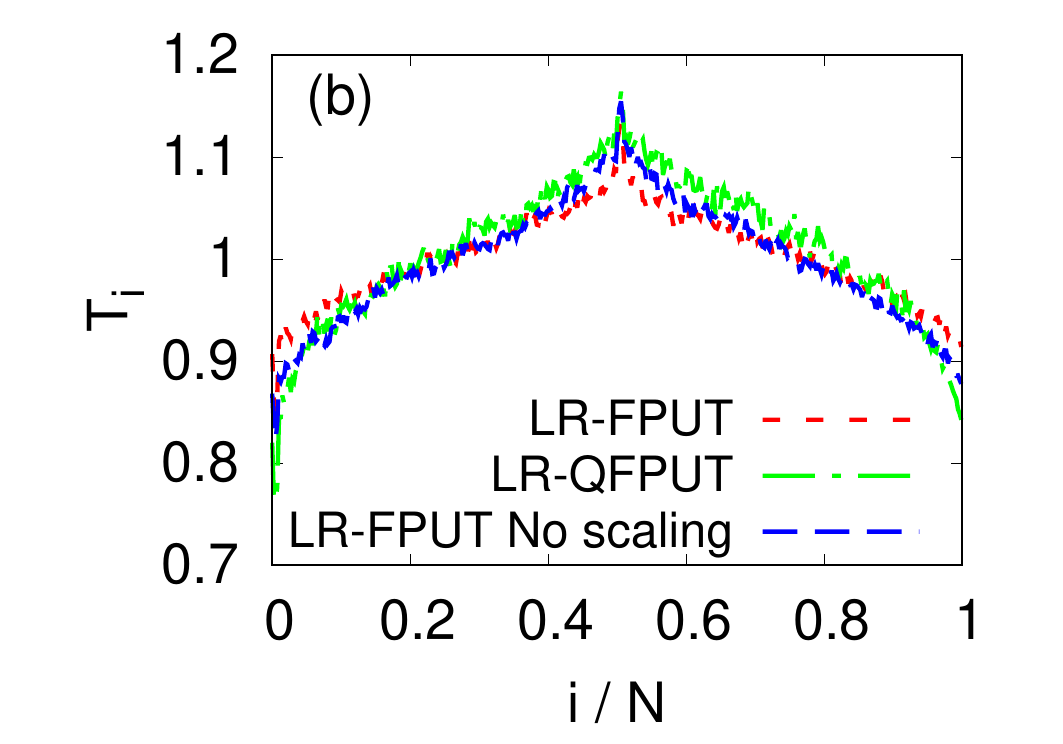}}
{\includegraphics[width=4.5cm]{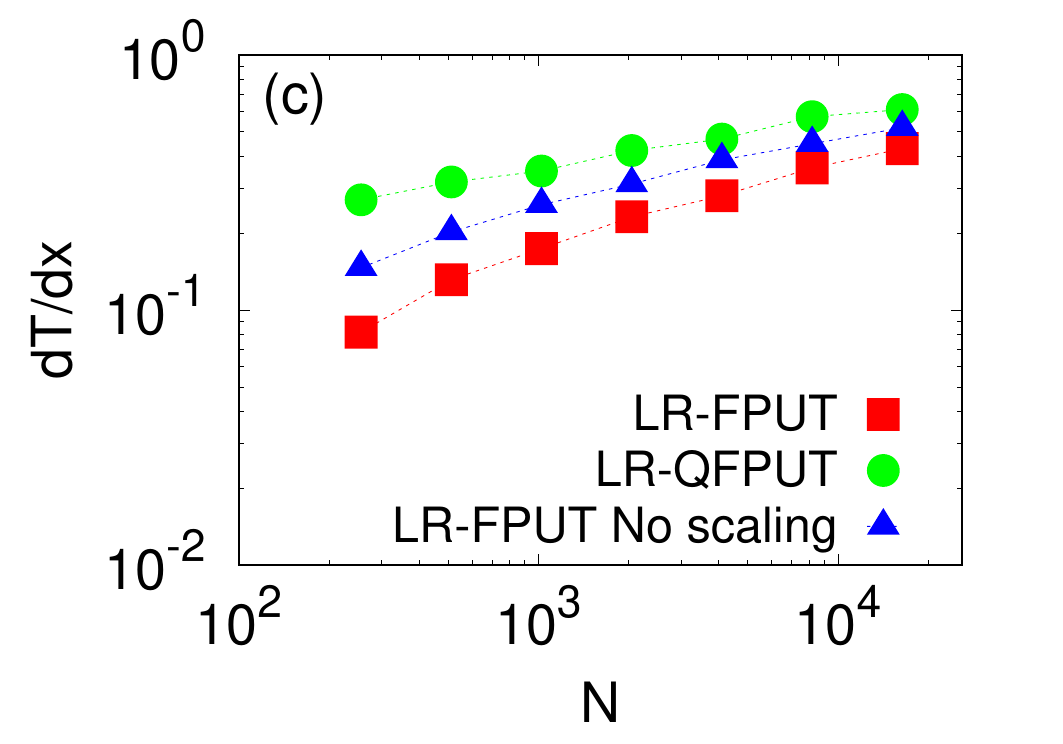}}
\caption{Results from RNEMD simulations: (a) Steady state total current obtained for the LR-FPUT model, the LR-QFPUT model, and the LR-FPUT model without $\widetilde{N}$ scaling. For all the three cases, the total current $J = jN$ scales linearly with $N$ at $\delta = 2$. The error bars here are smaller than the symbols used. (b) Temperature profiles for the three models with $N = 4096$. (c) The slope $dT/dx$ as a function of system size $N$. Here, the simulation parameters used are $\tau_s = 50$, $n = 40$, $n_s = 2$.}
\label{fig:rnemdcurr}
\end{figure}

Since the thermal flux is accurately known in RNEMD, we first look at its scaling in the steady state. The result for different system sizes, $128 \leq N \leq 16384$, is shown in Fig. \ref{fig:rnemdcurr}a. In this figure we have included the data for three models: LR-FPUT, LR-QFPUT, and LR-FPUT without the $\widetilde{N}$ scaling (as in Ref. \cite{wang2020thermal}). For all these models, we find the same linear scaling for the steady state total current $J \equiv jN \sim N$. The temperature profiles for these three models are very similar and the slopes of the temperature profile $\frac{dT}{dx}$ also have similar $N$ dependence (for large $N$), see Fig. \ref{fig:rnemdcurr}b,c. These results suggest that these three, apparently different, models have very similar scaling properties at $\delta=2$. Therefore, the suggestion in Ref. \cite{wang2020thermal} that there are strong finite-size effects due to the $\widetilde{N}$ scaling of the long-range potential and the use of boundary heat baths, leading to the large discrepancy in $\alpha$ values, seems very unlikely. We will provide more evidence in favor of this equivalence in Sec. \ref{DB}.

However, apart from the above-mentioned equivalence among the three models, we are not in a position to infer conclusively anything quantitative about the linear divergence of $J$ with size $N$ in Fig. \ref{fig:rnemdcurr}a. This is primarily because, for the same simulation parameters and with a large value of $\delta = 10$, we do not recover the expected short-ranged scaling relations, namely, $J \equiv jN \sim N^{\alpha}$ with $\alpha = 1/3$ \cite{dhar2008heat}. Using RNEMD simulations, the scaling exponent that we obtain for $J$ at $\delta=10$ is much higher (by a factor of $\sim 2-3$) than what is expected. Moreover, there is also a significant dependence of the scaling exponent $\alpha$ on the simulation parameters, such as $\tau_s$, number of particles in each slab $n$, and the particles that are swapped $n_s$ (see Fig. A.2a in Appendix). Surprisingly, we do not find such parameter dependence for $\delta=2$ (see Fig. A.2b in Appendix). Why we get such results with very reasonable parameter choices is beyond our present understanding. This makes us less confident about the reliability of the RNEMD method to compute transport properties in these systems.


We now discuss the issues related to the computation of the thermal conductivity $\kappa$ using the RNEMD simulations. One significant conceptual difference between the NEMD and the RNEMD methods lies in the manner in which $\kappa$ is computed. For NEMD, conductivity is usually computed as $\kappa = \frac{jN}{T_1 - T_2}$, whereas for RNEMD, the conductivity $\kappa = \frac{j}{dT/dx}$ has to be computed using the slope $dT/dx$ of the ``linear region'' of the temperature profile between the cold and the hot slab. It is easy to see that these two approaches lead to the same result for systems that obey Fourier's law and have a linear temperature profile. However, even with a small temperature difference between the hot and cold slabs (as in our case), the linear region is not straight-forward to ascertain for anomalous systems, which typically have nonlinear temperature profiles. This, in our opinion, could be a possible source of discrepancy in the thermal conductivity calculations. Indeed, when there is a larger temperature difference between the hot and the cold slabs, the temperature profiles are extremely nonlinear (see Fig. A.2c in Appendix), and thus, there is an inherent ambiguity in the computation of the slope, since $\frac{dT}{dx}$ changes continuously along such nonlinear temperature profiles. In fact, significant mismatches (by a factor of $\sim 5-10$) between RNEMD and NEMD slopes $dT/dx$ have been observed previously, see, for example, Supplementary Figure 4 in Ref. \cite{xu2014length}. Also, for RNEMD simulations, problems of nonlinear profiles and underprediction of transport coefficients have been reported in the context of other transport coefficients, such as the shear viscosity, even for a simple 3D Lennard-Jones fluid \cite{tenney2010limitations}. Under these circumstances, RNEMD yields unreliable measurements, and NEMD is known to be much more robust without any noticeable efficiency cost.

We believe that similar problematic issues are present in the RNEMD study reported in Ref.  \cite{wang2020thermal}, and the underprediction of the exponent $\alpha \approx 0.71$ reported therein has seemingly very little to do with $\widetilde{N}$ scaling of the long-ranged potential or the use of heat reservoirs in NEMD simulations. The three models discussed above are equivalent to each other with regards to their $\delta=2$ transport behavior, and this is also consistent with what we find from NEMD simulations (same equivalence seen in Figs. A.3 and A.4 in Appendix). Thus, given these conceptual and technical concerns, our general suggestion will be to exercise caution in comparing results from RNEMD with NEMD simulations for $1D$ models with non-Fourier transport properties.

\section{Energy relaxation}
\label{Erelax}
In order to better appreciate the unique transport behavior at $\delta=2$, we now look at the process of energy relaxation in an open LR-FPUT chain with fixed boundary conditions ($x_0 = x_{N+1} = 0$). For a system size of $N=512$ particles, we first thermalize our LR-FPUT chain by attaching a Langevin heat bath at temperature $T = 1$ to each of the oscillators. After thermalization has been achieved, we remove the heat baths except at the two ends for which we set $T = 0$ with friction coefficient $\zeta = 0.1$ (same as in Ref. \cite{piazza2003cooling}) to allow energy leakage through the boundaries. Thus the average temperature (and the total energy) of the system is expected to decrease as time progresses, and ideally should become zero at large times.
\begin{figure}[htb]
\centering
{\includegraphics[width=8cm]{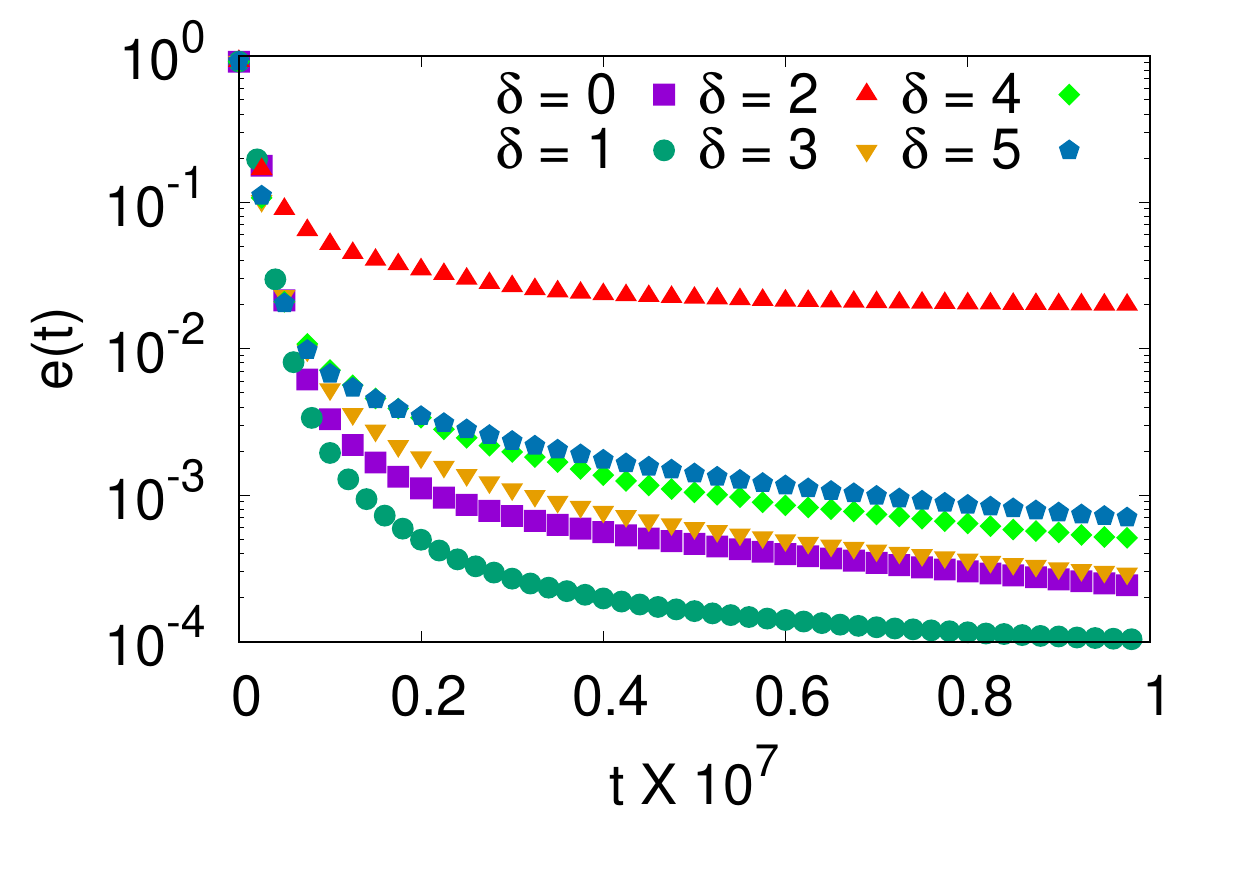}}
\caption{Energy density $e(t)$ with time $t$ in the LR-FPUT model for different values of $\delta$, with $N = 512$, initial temperature $T=1$, and boundary friction $\zeta = 0.1$. Here, $\delta=2$ shows relatively slower energy relaxation compared to other values of $\delta\neq2$.}
\label{fig:erelax}
\end{figure}

We monitor the decay of the energy density (average energy per oscillator) of the system $e(t)$ with time $t$, and the data is shown in Fig. \ref{fig:erelax} for different values of $\delta$. Even for a modest system size $N=512$, we find that the energy density of the system at late times does not decay to zero for $\delta = 2$ ($e(t) > 10^{-2}$ at $t = 10^7$), whereas for all other $\delta$ the energy density decays comparatively faster to relatively smaller values ($e(t) < 10^{-3}$ for $t = 10^7$). Thus, under these simulation conditions, the $\delta=2$ system remains trapped in a long-lived quasi-stationary {\it residual state} \cite{piazza2003cooling} and exhibits non-exponential energy relaxation, similar to what is observed in glassy systems \cite{tsironis1996slow}. While $e(t)$ for all $\delta \neq 2$ values show a decaying trend at large times, the energy decay profile for $\delta=2$ is comparatively very flat, and the system seems to be frozen in the residual state. This result demonstrates that under the above-mentioned simulation conditions $\delta = 2$ again stands out as a special point for its exceedingly slow energy relaxation, as compared to all other $\delta$ values.

\section{Breather Dynamics}
\label{DB}
\begin{figure*}[htb]
{\includegraphics[width=14.5cm]{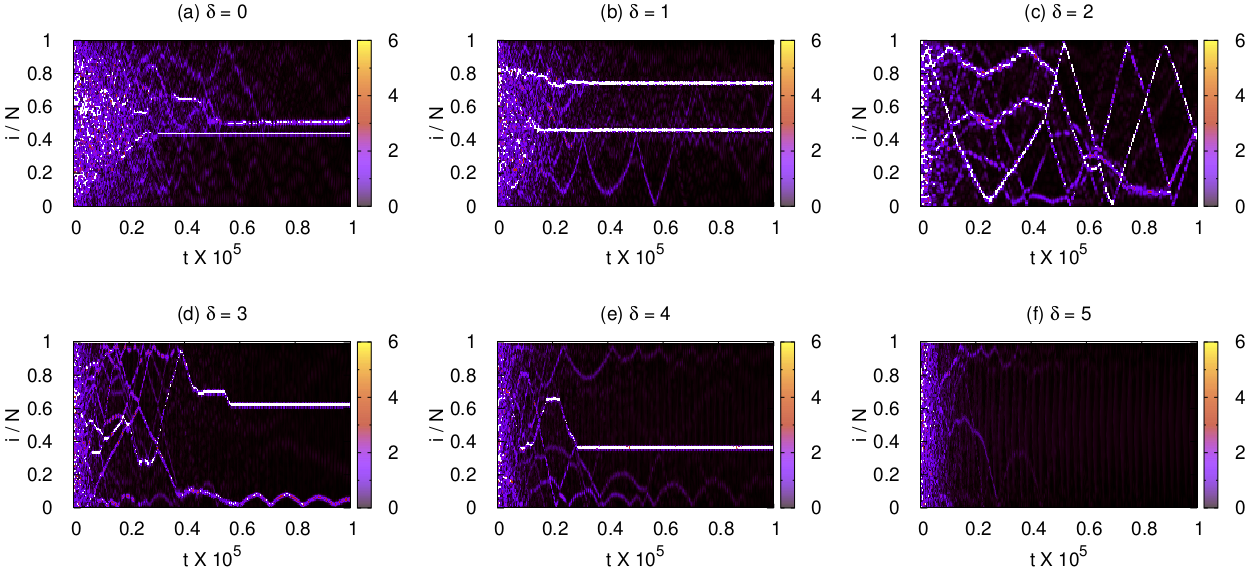}}
\caption{(a)--(f) Heat-maps of the LR-FPUT model for different values of $\delta$ ($0 \leq \delta \leq 5$) for $N=128$. The color bar represents the value of temperature $T_i$. Initial temperature is $T=3$ and boundary friction is set to $\zeta=1$. Highly mobile TDBs can be seen in (c) for $\delta=2$.}
\label{fig:map}
\end{figure*}

To probe the slow energy relaxation at the single oscillator level in the LR-FPUT model, we anticipate the presence of energy localization by spontaneously generated discrete breathers (DBs) \cite{tsironis1996slow,flach1998discrete,kivshar2003introduction,piazza2003cooling,flach2008discrete} that can prevent rapid relaxation. Discrete breathers are defined as intrinsic nonlinear localized modes of excitation with a time periodic structure, and have been observed in nonlinear discrete lattices under very generic conditions. Here, we examine DB dynamics following the boundary cooling experiment protocol adopted in Ref. \cite{bagchi2017energy}. As before, we first thermalize the open chain to a relatively high temperature, $T = 3$ by attaching Langevin heat baths to all the oscillators on the chain. After thermalization, we remove the heat baths from all the bulk sites and set the boundary heat baths to $T_1 = T_2 = 0$ (with friction set to $\zeta=1$). For a relatively small system, $N = 128$, we monitor the kinetic temperature, $T_i = v_i^2$, of each oscillator on the chain as a function of time. 

The local temperature $T_i(t)$ for different $\delta$ values is shown in the heat-maps of Fig. \ref{fig:map}(a)-(f). From Fig. \ref{fig:map}(c) for $\delta = 2$, we find that there are many bright lines criss-crossing the heat-map until very late times. These zigzag lines are {\it hot spots} of trapped (localized) energy that can travel from one end of the lattice to the other with seemingly negligible dissipation (energy loss). Careful visual inspection reveals that, when the DBs reach near the boundaries, they get reflected back toward the bulk of the lattice, and continue their motion unhindered along the reflected trajectory. In Fig. \ref{fig:map}, we find such highly mobile traveling discrete breathers (TDBs) only for $\delta = 2$, whereas, for all other values of $\delta \neq 2$ the DBs seem to be almost pinned with very low mobility at late times. This unique behavior of the DBs at $\delta=2$ leads us to believe that it is this virtually unhindered motion of the long-lived localized excitations that is responsible for the slow energy relaxation (in Fig. \ref{fig:erelax}) and the ballistic-like thermal transport in the LR-FPUT model at $\delta = 2$.

\begin{figure}[htb]
\centering
{\includegraphics[width=5cm]{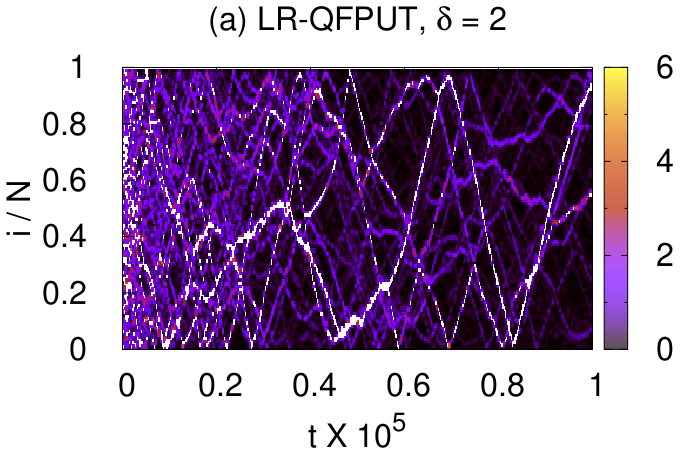}}\vskip0.3cm
{\includegraphics[width=5cm]{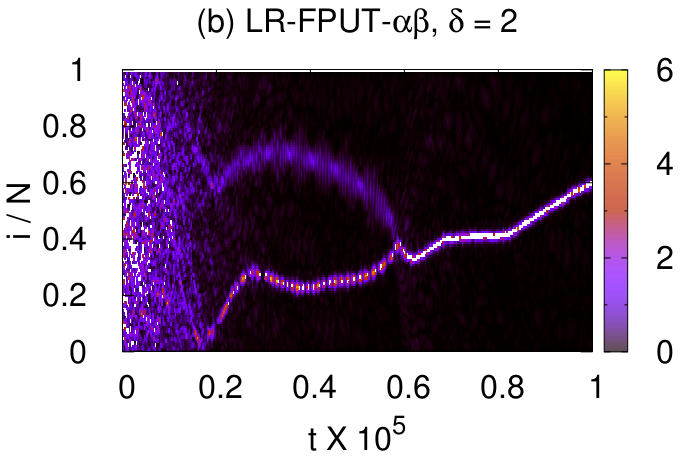}}
\caption{Prevalence of highly mobile TDBs in (a) the LR-QFPUT model, and its absence in (b) the LR-FPUT-$\alpha\beta$ model, at large times, for $\delta=2$. All the simulation parameters are the same as in Fig. \ref{fig:map}.}
\label{fig:map3}
\end{figure}

As a further check of this TDB hypothesis, we perform the same numerical experiment with the LR-QFPUT model (Eq. \ref{eq:lrqfput}). The transport properties of LR-QFPUT are very similar to the LR-FPUT model: one obtains a non-monotonic $\kappa$ $vs$ $\delta$ with maximum at $\delta = 2$, linear $\kappa \sim N$ dependence, and an almost flat temperature profile at $\delta = 2$ (see Fig. A.4a-c in Appendix). For this case too, inspection of its heat-map, Fig. \ref{fig:map3}a, reveals that the DBs are highly mobile at $\delta = 2$.
Since there is very little impedance to the motion of these TDBs (negligible scattering), this could be a plausible explanation for the almost flat temperature profiles observed at $\delta=2$.
This also goes to prove that the short-ranged harmonic term in the LR-FPUT model (Eq. \ref{eq:lrfput}) has no role to play in the ballistic-like heat transport observed at $\delta=2$. 

Thus, considering results from all the variant models from this work and previous works \cite{wang2020thermal,iubini2018heat,bagchi2017thermal}, we infer that the ballistic-like transport is a robust property of these long-ranged models, and owes its origin to the highly mobile TDBs that emerge at the special point $\delta=2$. It is relevant to mention here that TDBs have also been observed and studied previously, even for the nearest-neighbor Fermi-Pasta-Ulam-Tsingou model  \cite{sandusky1992stability,bickham1993stationary,flach1994movability}.

We can make the TDB argument even stronger by asking if the opposite scenario is possible, i.e., can we obtain non-ballistic transport at $\delta=2$ by ``killing off'' the TDBs? To check if this is true, we consider another long-ranged model with additional nearest-neighbor cubic interactions, which, we suspect, will create dissipation of the TDBs and make them sufficiently immobile at long times. Following the standard nomenclature, we refer to this model as the long-ranged Fermi-Pasta-Ulam-Tsingou-$\alpha\beta$ model (LR-FPUT-$\alpha\beta$); the coefficient $\alpha$ here is set to unity in our simulations and should not be confused with the exponent $\alpha$ in $\kappa \sim N^\alpha$. The Hamiltonian for this case is
\begin{multline}
\textbf{LR-FPUT-}\alpha\beta: \\ \mathcal{H}_3 = \sum_{i}\frac 12 p_i^2 + \sum_{i}\frac k2 (x_{i+1}-x_i)^2 + \frac {\alpha}3 \sum_{i} (x_{i+1}-x_i)^3 \\ + \frac{\beta}{4\widetilde{N}} \sum_{i,j>i} \frac{(x_j-x_i)^4}{d_{ij}^\delta}.
\end{multline}

The result for the LR-FPUT-$\alpha\beta$ model is shown in Fig. \ref{fig:map3}b and it is evident that the DBs are much more immobile in this case, compared to the LR-FPUT (Fig. \ref{fig:map}c) and the LR-QFPUT (Fig. \ref{fig:map3}a) models for $\delta=2$. As anticipated, we have checked that ballistic-like transport properties are not observed for the LR-FPUT-$\alpha\beta$ case (see Fig. A.5a-b in the Appendix). All these results strongly suggest that these TDBs are responsible for the emergence of the ballistic-like transport seen for $\delta = 2$ in the LR-FPUT model (and also in LR-QFPUT). Note that a similar argument has been suggested in Ref. \cite{wang2020thermal} for the LR-FPUT model without $\widetilde{N}$ scaling. Thus, we again find that these three models -- LR-FPUT model with and without $\widetilde{N}$ scaling, and the LR-QFPUT model -- are equivalent, and exhibit the same TDB phenomenon.

Why we observe high breather mobility at $\delta=2$ in the LR-FPUT model is an interesting question that needs further investigation. Incidentally, a similar question has been recently addressed using a long-ranged model that the authors refer to as the pairwise interaction symmetric lattice (PISL) model \cite{doi2016symmetric}. It has been shown that a PISL system can support smooth propagation of discrete breathers at constant velocity, and this depends crucially on the symmetry properties of the potential function. It seems plausible that such a reasoning could also be true for the LR-FPUT model.

Thus, to reiterate, from Fig. \ref{fig:map} we find that as one increases $\delta$ from zero, the DBs transform from being predominantly immobile to highly mobile at the special point $\delta=2$, and for $\delta > 2$ they become immobile again. This interesting ``pinning $\to$ depinning $\to$ pinning'' phenomenon exhibited by the DBs, as we increase $\delta$ from zero, seems to be responsible for the puzzling transport properties observed in the LR-FPUT model reported in Ref. \cite{bagchi2017thermal}. 

Note that the exact evolution of the DB dynamics at large times will depend on several factors, such as fixed or free boundaries, the friction coefficient $\zeta$, the initial energy density of the thermalized state \cite{piazza2003cooling}, and so on. In this section, we have used the simulation conditions similar to that used for the NEMD simulations in Ref. \cite{bagchi2017thermal} with $\zeta = 1$. For boundaries with a lower friction $\zeta = 0.1$, as used in the previous section Sec. \ref{Erelax}, the DBs generated are found to be weaker and much less prominent at late times (see Fig. A.6a--f in Appendix). However, long-lived TDBs are still found to be present at $\delta = 2$ for the LR-FPUT model (see Fig. A.6c in Appendix) and, therefore, they seem to be quite robust with respect to simulation parameters. A detailed study of the DB properties in the LR-FPUT model, such as their profiles, lifetime, frequencies, stability, mobility, and their dependence on the system parameters, such as system size, temperature, interaction strengths, and boundary conditions, should be of interest in their own right.

\section{Fourier behavior}
\label{pinning}
From the previous section, Sec. \ref{DB}, we find that the ballistic-like heat transport at $\delta = 2$ can be attributed to the highly mobile TDBs, and by adding a short-ranged anharmonic interaction term in the Hamiltonian one obtains super-diffusive transport. In this section, we want to investigate the possibility of observing diffusive transport (Fourier behavior) in the LR-FPUT at $\delta = 2$. From our general understanding of heat transport in short-ranged interacting systems, one definitive way of obtaining diffusive transport is by breaking total momentum conservation. However, in Ref. \cite{tamaki2020energy}, it has been suggested that for long-ranged systems, even momentum non-conservation may lead to anomalous (non-Fourier) behavior, contrary to what is typically observed in short-ranged systems. Note that this prediction is obtained from the analytical study of a quadratic long-ranged model with harmonic pinning.

To explore the effect of pinning potentials in the LR-FPUT model, we augment the LR-FPUT Hamiltonian (Eq. \ref{eq:lrfput}) by adding a pinning potential of the form $U(x) = \frac up x^p$. Here, the constant $u > 0$ is the strength of the pinning potential and the exponent $p > 0$. Explicitly written, the Hamiltonian for the pinned case is given as

\begin{multline}
\textbf{Pinned LR-FPUT}: \\ \mathcal{H}_4 = \sum_{i}\frac 12 p_i^2 + \sum_{i}\frac k2 (x_{i+1}-x_i)^2 + \frac up \sum_{i} x_i^p \\ + \frac{\beta}{4\widetilde{N}} \sum_{i,j>i} \frac{(x_j-x_i)^4}{d_{ij}^\delta}.
\end{multline}

A reliable and computationally economical method for ascertaining the nature of energy transport is by computing equilibrium spatio-temporal excess energy correlation function $\rho_E(r,t)$, defined as
\begin{equation}
 \rho_E(r,t) = \frac{\langle \Delta E_j(t)  \Delta E_i(0) \rangle}{\langle  \Delta E_i(0)  \Delta E_i(0) \rangle} + \frac{1}{N_b - 1},
 \label{pinned}
\end{equation}
for a microcanonical system. Here, the lattice is coarse-grained into $N_b = N/b$ bins, each with $b$ particles, and $r = (i-j)b$; see \cite{bagchi2017thermal, bagchi2017energy} and references therein for more details. For Fourier (diffusive) transport, it is well known that $\rho_E(r,t)$ has to be a Gaussian distribution, and therefore $\rho_E(r,t)$ at different times can be collapsed by scaling as $t^\gamma\rho_E(r/t^\gamma,t)$, with $\gamma = 1/2$.

\begin{figure}[htb]
\centering
{\includegraphics[width=7cm]{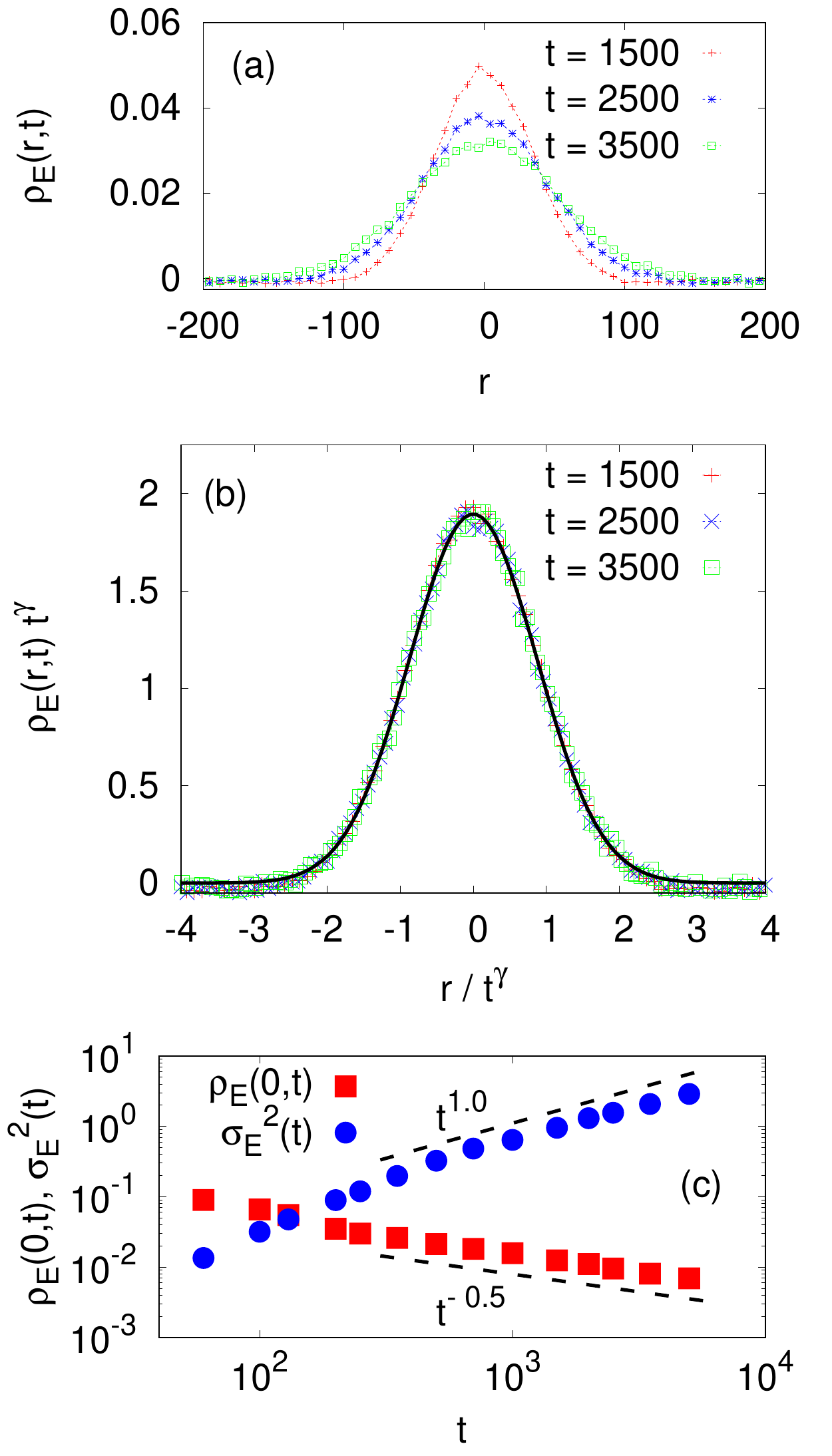}}
\caption{Equilibrium simulation results for the LR-FPUT model with quartic pinning potential: (a) excess energy correlation function $\rho_E(r,t)$ at different times $t = 1500,2500,3500$; (b) scaling collapse of the data in (a) for $\gamma=1/2$, plotted alongside a Gaussian distribution (continuous line in black); (c) log--log plot of the peak height and the MSD of $\rho_E(r,t)$ with $t$. The MSD is scaled by the system size ($N = 4096$) in order to display both sets of data in the same plot.}
\label{fig:spread}
\end{figure}

The results for the pinned LR-FPUT model, Eq. (\ref{pinned}), with a quartic pinning potential $U(x) = \frac u4 x^4$ are presented in Fig. \ref{fig:spread}. The function $\rho_E(r,t)$ at different times $t$ is shown in Fig. \ref{fig:spread}a, and at least numerically looks very similar to a Gaussian distribution. The data points at different times $t$ can be collapsed onto a single curve using the scaling mentioned above (Fig. \ref{fig:spread}b) and the agreement with a Gaussian distribution is excellent. Consistently, the distribution peak height $\rho_E(0,t)$ decays with time as $\rho_E(0,t) \sim t^{-1/2}$ and the mean square deviation (MSD) of the distribution scales as $\sigma^2_E(t) = \sum r^2 \rho_E(r,t) \sim t$, as shown in  Fig. \ref{fig:spread}c. We have obtained strong indications of diffusive transport from NEMD simulations as well, such as $\kappa \sim N^{0}$ for large $N$ and linear temperature profiles (see Figs. A.7a,b in Appendix). It is also found that, as we increase $u$, $\delta=2$ gradually ceases to become a special point with maximum $\kappa$ and for larger values of $u$ conductivity $\kappa$ increases monotonically with $\delta$ (see Fig. A.7c in Appendix). These results strongly indicate toward diffusive propagation of energy fluctuations in the LR-FPUT model with quartic pinning potential, as one would expect also for short-ranged interacting systems with broken momentum conservation.

\begin{figure}[htb]
\centering
 {\includegraphics[width=5cm]{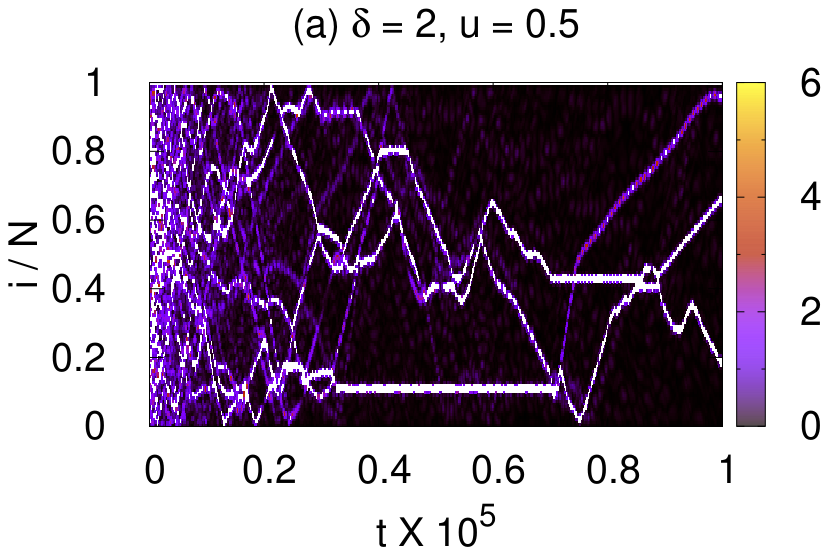}}\vskip0.35cm
 {\includegraphics[width=5cm]{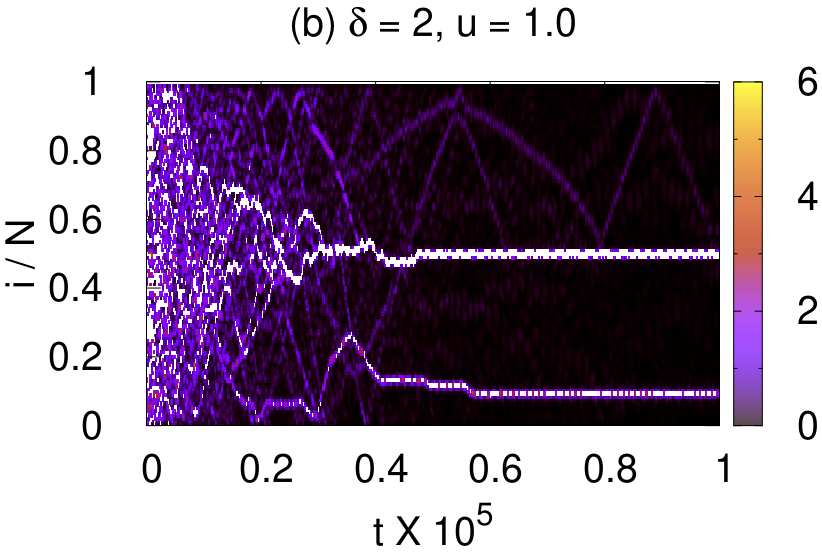}}\vskip0.35cm
 {\includegraphics[width=5cm]{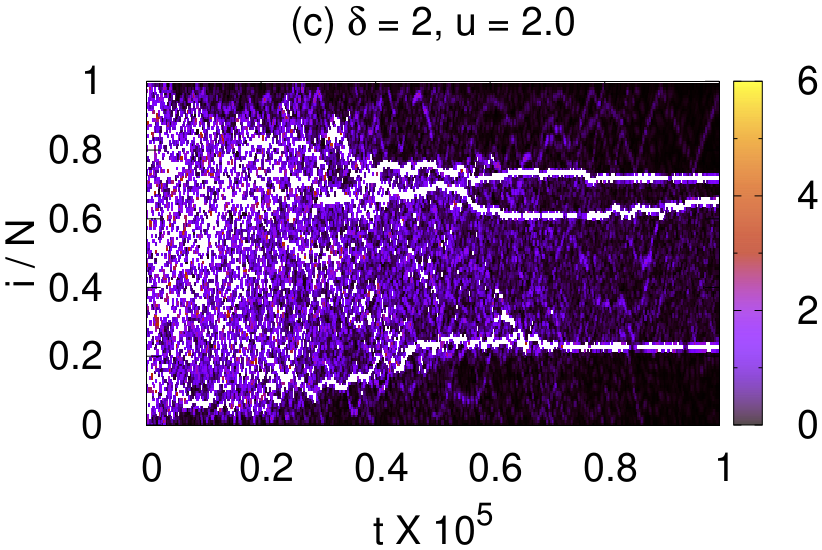}}
\caption{DBs in the LR-FPUT model with quartic pinning for different pinning strengths: (a) $u = 0.5$, (b) $u = 1$, and (c) $ u = 2$. The DBs lose their mobility as $u$ is increased.}
\label{fig:DBpin}
\end{figure}

To understand the emergence of Fourier behavior, we again look at the dynamics of the discrete breathers, as shown in Figs. \ref{fig:DBpin}a-c for $u = 0.5,1$, and $2$. From Fig. \ref{fig:DBpin}, first, it can be clearly seen that the TDBs progressively lose their mobility as $u$ is increased: there are more straight horizontal lines that emerge and the zigzag lines start to disappear in the heat-maps (compare this with Fig. \ref{fig:map}c where $u=0$). Second, as $u$ is increased, DBs become comparatively more pronounced: the heat-map becomes brighter in color at larger $t$, implying higher energy localization in the bulk of the system. This increased energy trapping leads to the slow-down of energy propagation, from ballistic to normal diffusion when $u \neq 0$.

Note that with harmonic pinning potential ($p = 2$) Fourier behavior is not observed, at least for the system parameters that we have studied. Our simulations  suggest ballistic transport even with a moderately strong harmonic pinning strength $u$ (see Fig. A.8 in Appendix). In general, harmonic pinning is often known to produce subtle features, even in short-ranged systems, such as the integrable $1D$ Toda lattice \cite{lebowitz2018ballistic,di2018transport,dhar2019transport}. For the Toda lattice, it has been recently shown that the effect of harmonic pinning is ``drastically smaller'' than quartic pinning potential. A crossover from ballistic to diffusive transport is also reported for the Toda lattice at very large $N$. Although, we cannot entirely rule out the possibility of a crossover to a diffusive regime for extremely large system sizes (and possibly well beyond our present computational capability), it is important to note from Ref. \cite{doi2016symmetric} that the nature of the TDBs seems to remain unaffected in the presence of harmonic pinning. In this context, an important $1D$ long-ranged system that should also be mentioned is the integrable Calogero model which is known to preserve its integrability even in the presence of a harmonic trap, and supports solitons in the $N \to \infty$ limit \cite{abanov2011soliton}.

\section{Summary and final remarks}
\label{conclusion}
In summary, we have studied in detail the heat transport problem in $1D$ anharmonic oscillator models with long-ranged interactions, such as the LR-FPUT model. We have looked at different scenarios, namely, with and without the $\widetilde{N}$ scaling, without the short-ranged harmonic potential, with additional short-ranged anharmonic potentials, and in the presence of pinning potentials.
Our simulations show that the transport properties of the LR-FPUT model, with and without the $\widetilde{N}$ scaling, are essentially equivalent to each other in their transport behavior. We have also pointed out a few conceptual and technical issues with the RNEMD method that warrant more careful scrutiny regarding its applicability in these anomalous systems.

To emphasize the uniqueness of the $\delta = 2$ transport, we look at the energy relaxation process in the LR-FPUT, and find that under certain conditions energy relaxation can be much slower for $\delta = 2$ compared to other $\delta$ values. Next, we demonstrate that the $\delta=2$ LR-FPUT model supports traveling discrete breathers that can propagate through the system with negligible energy loss until very late times. By comparing the DBs for different values of $\delta$, we attribute the unique transport property at $\delta = 2$ to the depinning phenomenon of the DBs that makes them highly mobile. This certainly seems to be the case for the LR-FPUT model (with and without $\widetilde{N}$ scaling) as well as for the LR-QFPUT model. As a further vindication of this argument, we show that these highly mobile TDBs are absent in the LR-FPUT-$\alpha\beta$ model which does not exhibit the peculiar ballistic-like transport at $\delta = 2$. From all these consistent results, the ballistic-like behavior at $\delta=2$ for the LR-FPUT model seems to be a real physical property and not merely a finite-size effect, although the latter possibility cannot be completely ruled out. 

Finally, we study the emergence of Fourier behavior in the LR-FPUT model at $\delta=2$. We break total momentum conservation by adding a quartic pinning potential and find that heat transport becomes diffusive. The emergence of this diffusive regime is attributed to the immobilization to the TDBs, and these immobile DBs are also more pronounced for strong quartic pinning potentials. This creates substantial impedance to the propagation of the DBs and slows down energy transport from a ballistic to a diffusive regime. However, with harmonic pinning we have not observed such a diffusive regime in our simulations.

Thus, by studying the properties of the discrete breathers, we can consistently explain the emergence of ballistic-like transport in the LR-FPUT and LR-QFPUT models, the super-diffusion observed in the LR-FPUT-$\alpha\beta$ model, and the diffusion in the quartic pinned case, observed at $\delta = 2$.

To make our understanding more quantitative about the origin of the highly mobile TDBs at $\delta = 2$, it might be instructive to study the depinning phenomenon in terms of the Peierls-Nabarro (PN) potential barrier. The PN barrier can be thought of as the minimum energy required to translate a DB by one lattice site. The PN barrier has been studied earlier for traveling DBs in the short-ranged FPUT model and other nonlinear lattices \cite{dauxois1993energy,aubry1998mobility,mackay2002effective,martinez2003dissipative,johansson2015breather}. We suspect that the PN potential barrier will perhaps be a non-monotonic function of $\delta$, with its minimum at $\delta = 2$, resulting in the enhanced mobility of the DBs.

Based on all these results, one can also speculate about a few plausible features of the LR-FPUT model, and attempt to establish possible connections with some related recent works. The yet unsettled question of quasi-integrable dynamics \cite{bagchi2017thermal} or weaker non-integrability \cite{wang2020thermal}, at $\delta=2$, might have an answer in terms of quasi-conservation law and adiabatic invariant (AI) \cite{gardner1959adiabatic} that have been proposed recently to explain slow energy relaxation (thermalization) in the discrete nonlinear Schrodinger (DNLS) equation \cite{iubini2019dynamical}. Similar to our case, the slow relaxation in the DNLS model has also been attributed to discrete breather dynamics. The presence of AI is thought to be responsible for the existence of small yet non-zero Lyapunov exponents \cite{iubini2021chaos}, although the largest Lyanpunov exponent $\lambda_{max}$ is found to remain insensitive to the DB solutions \cite{mithun2018weakly}. In contrast, for the LR-FPUT model, $\lambda_{max}$ is clearly affected by the presence of the traveling DBs, and one obtains a minimum for $\lambda_{max}$ (weak chaos) at $\delta=2$; see the measurement of $\lambda_{max}$ as a function of $\delta$ in Refs.  \cite{bagchi2017thermal,di2019equilibrium,wang2020thermal}. Thus, although AIs are not exact conservation laws, in some cases they may have effects similar to quasi-integrable dynamics, e.g., when there are long-lived traveling discrete breathers in non-integrable systems. Also, the phase space dynamics for the LR-FPUT could be an interplay of chaotic trajectories and regular orbits, and their relative proportions depend, possibly in a complicated way, on the range parameter $\delta$; see Ref. \cite{bachelard2008abundance} for a similar description of regular orbits in the paradigmatic Hamiltonian mean-field model. However, it is also possible for a system to remain well beyond the KAM/Nekhoroshev regime and yet behave as a weakly chaotic system; see discussions on stochastically perturbed integrable systems in Ref. \cite{lam2014stochastic}. Whether these arguments are at all true for the LR-FPUT model is not known yet, but it seems worthwhile to examine these aspects for a better understanding of many key concepts, such as ergodicity, equipartition, thermalization, localization, and transport properties in the LR-FPUT and related long-ranged models.

To conclude, we have used various simulation techniques (RNEMD, NEMD, equilibrium methods) and different probing methods (energy relaxation, scaling of current and conductivity, temperature profiles, energy correlations, discrete breather dynamics) to unravel the heat transport properties of a class of $1D$ anharmonic oscillators with long-ranged interactions. The results presented here demonstrate the rich transport behavior observed in these systems. This is a relatively new area of inquiry, and a lot remains to be explored and understood properly. Hopefully, our results will motivate further research in this direction, both from the standpoint of fundamental science and potential technological applications.


\bibliographystyle{apsrev4-2}
\bibliography{References}



\onecolumngrid

\pagebreak

\section*{Appendix: Heat transport in long-ranged anharmonic oscillator models}

\appendix
\renewcommand\thefigure{A.\arabic{figure}}
\setcounter{figure}{0}    

\vskip1cm

\begin{figure}[htb]
{\includegraphics[width=16cm]{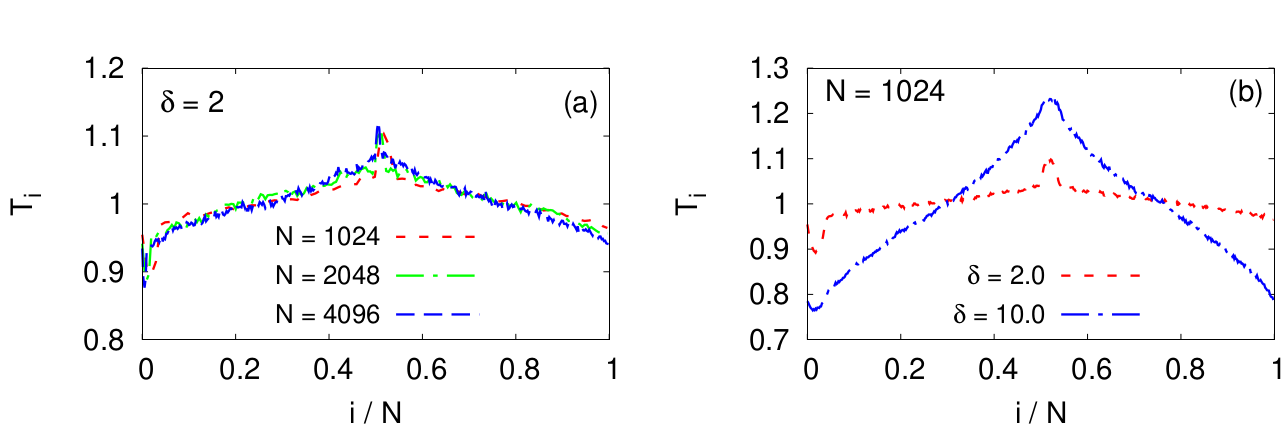}}
\caption{RNEMD temperature profiles for the LR-FPUT model: (a) different system sizes $N$ at $\delta=2$, and (b) comparison between the temperature profiles for $\delta=2$ and $\delta=10$ with $N = 1024$.}
\label{fig:rnemdTP1}
\end{figure}

\vskip1cm

\begin{figure}[htb]
\centering
 {\includegraphics[width=12cm]{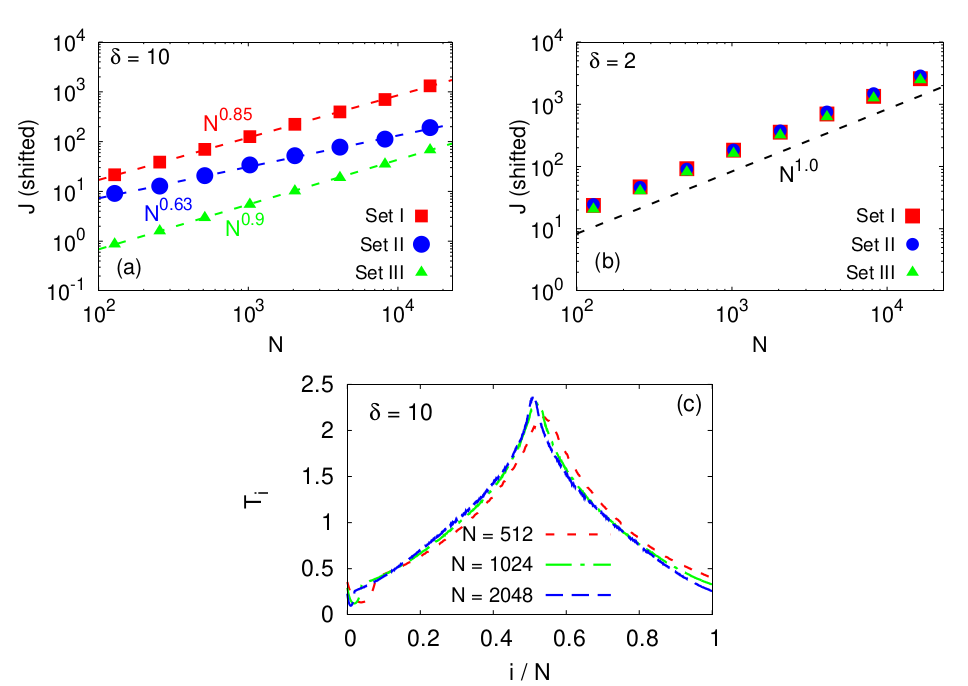}}
\caption{RNEMD results for the steady state total energy current $J = jN$ in the LR-FPUT model for (a) $\delta = 10$ and (b) $\delta = 2$, for three different parameter sets: Set I ($\tau_s = 50$, $n = 20$, $n_s = 2$), Set II ($\tau_s = 50$, $n = 40$, $n_s = 15$), and Set III ($\tau_s = 10$, $n = 40$, $n_s = 2$). For $\delta = 10$, the scaling exponent $\alpha$ is not robust, unlike $\delta = 2$, with respect to different simulation parameter sets. The data for the different sets have been shifted along the $y-$axis (by a multiplicative factor) for better visibility. (c) Nonlinear temperature profiles for $\delta=10$ with a large temperature difference between the cold slab and the hot slab (parameters are the same as for Set II).}
\label{fig:RNEMDa10}
\end{figure}

\begin{figure}[htb]
\centering
{\includegraphics[width=12cm]{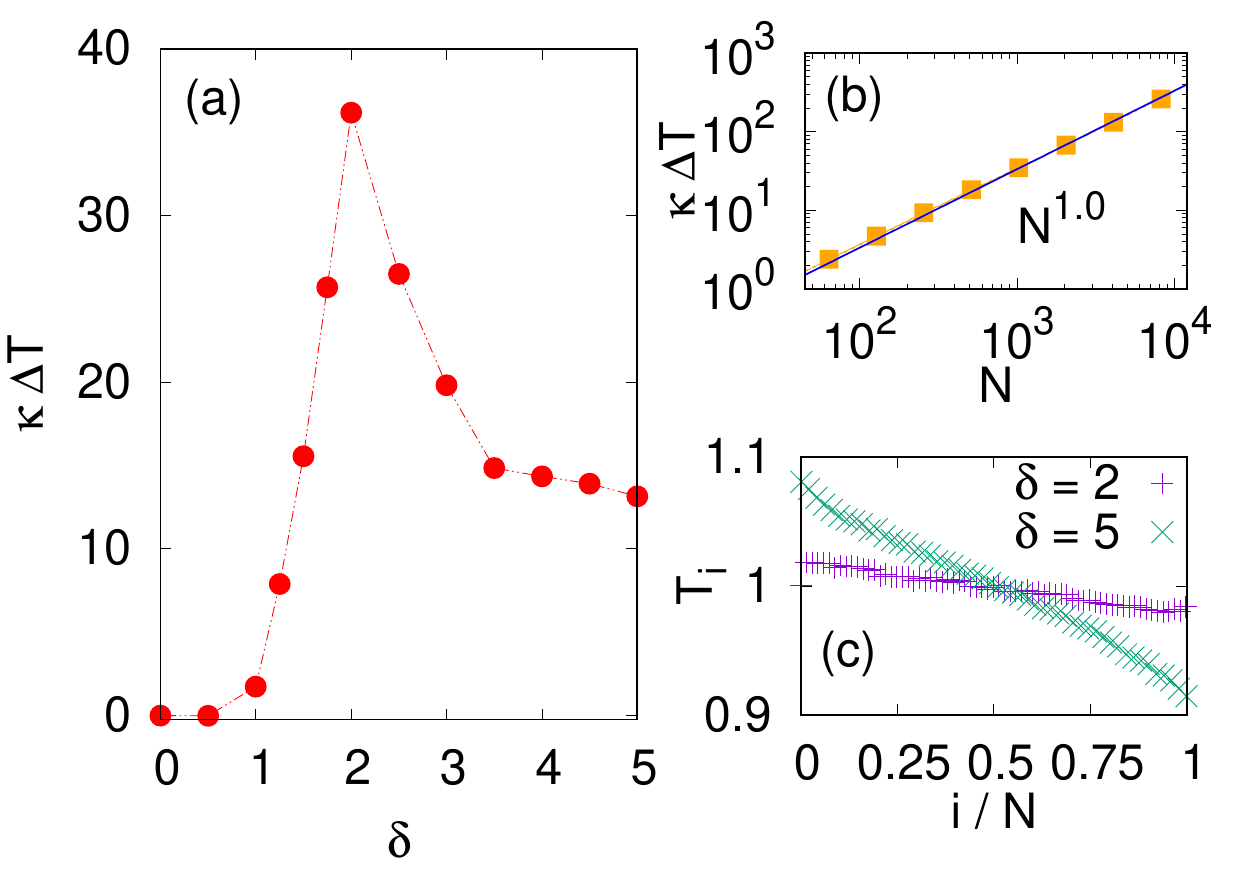}}
\caption{Transport properties of the LR-FPUT model without the $\widetilde{N}$ scaling, obtained using NEMD simulations. (a) Thermal conductivity $\kappa$ as a function of $\delta$ for $N = 1024$. (b) Conductivity $\kappa$ with system size $N$ for $\delta=2$ shows a linear divergence. (c) Temperature profile for $N=1024$ with $\delta=2$ and $\delta=5$.}
\label{fig:lrfputN}
\end{figure}
%

\begin{figure}[htb]
\centering
{\includegraphics[width=12cm]{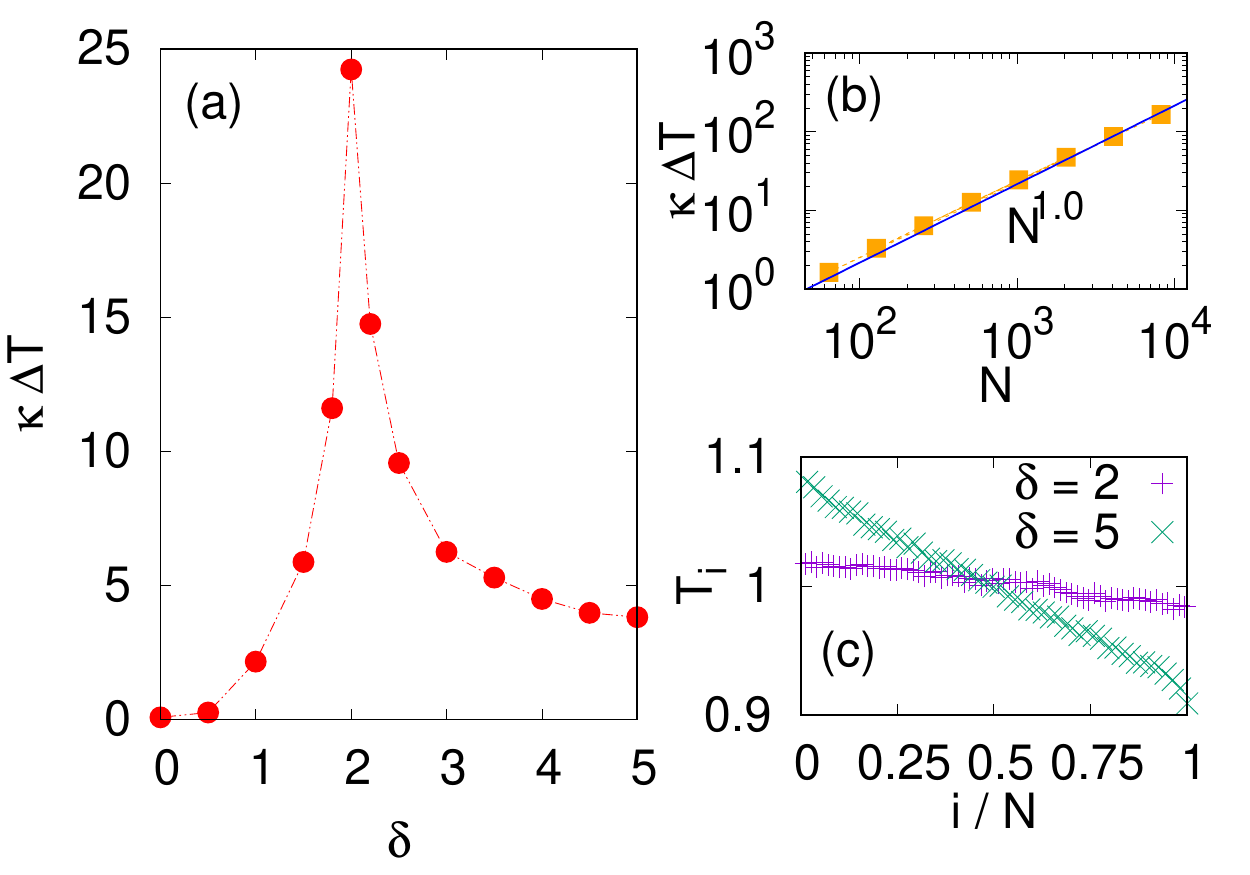}}
\caption{Transport properties of the LR-QFPUT model obtained using NEMD simulations. All parameters are the same as in Fig. \ref{fig:lrfputN}.}
\label{fig:lrqfput}
\end{figure}
%

\begin{figure}[htb]
\centering
 {\includegraphics[width=18cm]{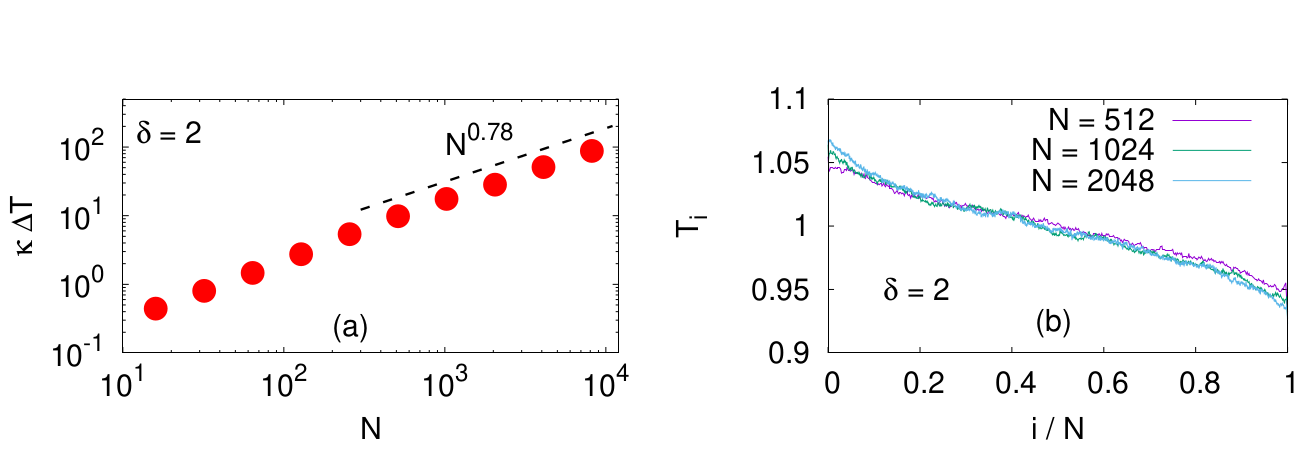}}
\caption{Super-diffusive transport properties of the LR-FPUT-$\alpha\beta$ model obtained using NEMD simulations: (a) Conductivity $\kappa$ with system size $N$ shows a sub-linear divergence $\kappa \sim N^{0.78}$ for $\delta=2$. (b) Temperature profiles $T_i$ for $N = 512, 1024$, and $2048$ are nonlinear, with an appreciable nonzero gradient for $\delta=2$.}
\label{fig:lr-fput-ab}
\end{figure}
%

\begin{figure*}[htb]
{\includegraphics[width=17.5cm]{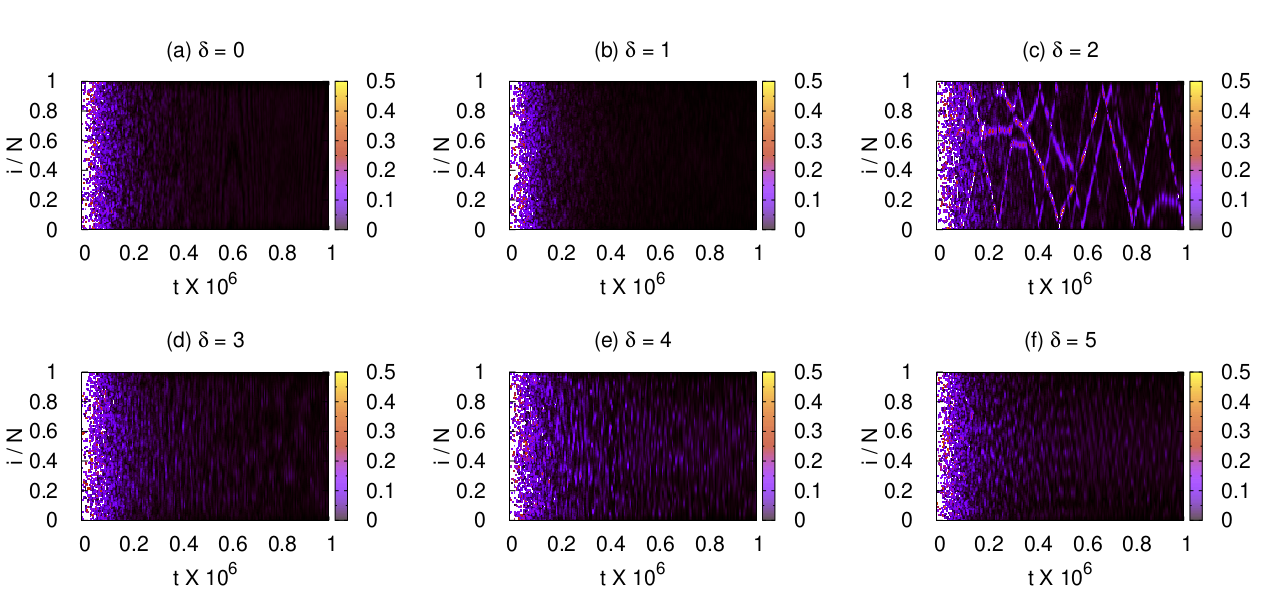}}
\caption{(a)-(f) Heat-maps for low boundary friction $\zeta=0.1$ and low initial temperature $T = 1$ in the LR-FPUT model, with $0 \leq \delta \leq 5$. For this case, the DBs are not observed at large times for $\delta \neq 2$, but a traveling DBs are observed for $\delta = 2$ and can slow down energy relaxation, as shown in Fig. 2 (main text). Here $N=128$.}
\label{fig:DB_lowg}
\end{figure*}

\begin{figure}[htb]
{\includegraphics[width=14cm]{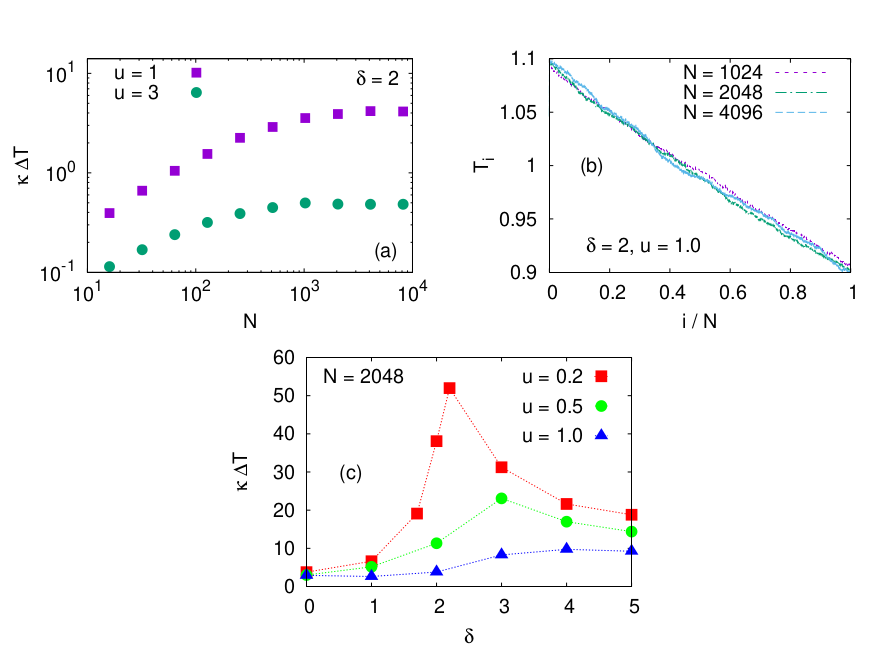}}
\caption{NEMD simulation results for the LR-FPUT model with quartic pinning: (a) Conductivity $\kappa$ saturates to a finite value at large $N$ for $\delta = 2$, and pinning strengths $u=1$ and $u = 3$. (b) Particle temperature $T_i$ forms a linear profile connecting the two bath temperatures $T_1 = 1.1$, and $T_2 = 0.9$. Thus, both (a) and (b) indicate a diffusive transport regime. (c) Conductivity $\kappa$ as a function of the interaction range parameter $\delta$ for different pinning strengths, $u = 0.2, 0.5$, and $1.0$, with $N=2048$.}
\label{fig:QuarticPin_JL}
\end{figure}

\begin{figure*}[htb]
\hskip5cm
{\includegraphics[width=18cm]{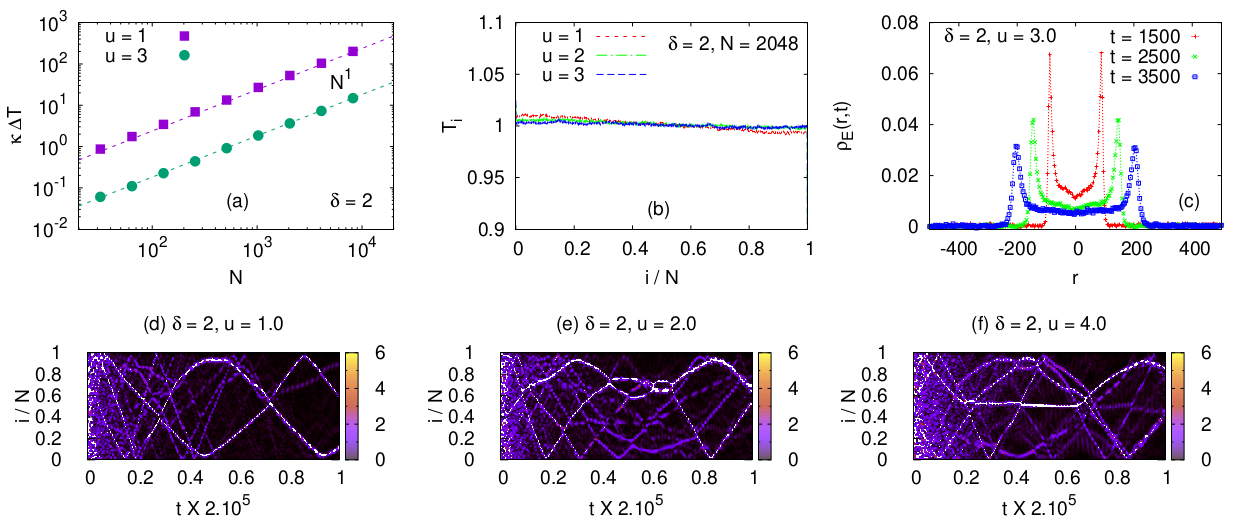}}
\caption{The LR-FPUT model with harmonic pinning at $\delta=2$: (a) Conductivity scales linearly with N for $u = 1$ and $u=3$. (b) Almost flat temperature profiles for different values of $u$ and $N=2048$. Boundary temperatures are $T_1 = 1.1$ and $T_2 = 0.9$. The temperature profile seems to get slightly flatter with increasing $u$. (c) Equilibrium excess energy correlation functions are not Gaussian. Here $N=4096$. Figs. (a)-(c) suggest a ballistic-like heat transport. (d)-(f) Heat-maps for $u = 1,2$, and $4$ show the presence of traveling discrete breathers ($N=128$).}
\label{fig:HarmPin_JL}
\end{figure*}

\end{document}